\documentclass[twocolumn]{aastex701}
\usepackage{amsmath,amssymb}

\begin{document}

\title{A Possible Hidden Hot Halo Component Revealed by Joint X-ray and Sunyaev--Zel'dovich Observations}

\author[orcid=0000-0001-6239-3821,sname='Li']{Jiang-Tao Li}
\affiliation{Purple Mountain Observatory, Chinese Academy of Sciences, 10 Yuanhua Road, Nanjing 210023, People’s Republic of China}
\affiliation{Department of Astronomy, University of Michigan, 311 West Hall, 1085 S. University Ave, Ann Arbor, MI 48109-1107, U.S.A.}
\email[show]{pandataotao@gmail.com}  

\author[orcid=0000-0001-7900-4204,sname='Huang']{Rui Huang}
\affiliation{Department of Astronomy, University of Michigan, 311 West Hall, 1085 S. University Ave, Ann Arbor, MI 48109-1107, U.S.A.}
\affiliation{Department of Astronomy, Tsinghua University, Beijing 100084, People’s Republic of China}
\email{huangrui@umich.edu}

\author[orcid=0000-0001-6276-9526,sname='Bregman']{Joel N. Bregman}
\affiliation{Department of Astronomy, University of Michigan, 311 West Hall, 1085 S. University Ave, Ann Arbor, MI 48109-1107, U.S.A.}
\email{jbregman@umich.edu}

\author[orcid=0000-0002-9279-4041,sname='Wang']{Q. Daniel Wang}
\affiliation{Department of Astronomy, University of Massachusetts, Amherst, MA 01003, U.S.A.}
\email{wqd@umass.edu}

\author[orcid=0000-0002-6653-8490,sname='Pratt']{Cameron Pratt}
\affiliation{Department of Astronomy, University of Michigan, 311 West Hall, 1085 S. University Ave, Ann Arbor, MI 48109-1107, U.S.A.}
\email{campratt@umich.edu}

\begin{abstract}

Massive galaxies are expected to host extended halos of hot gas that contain a substantial fraction of their baryons and feedback energy. X-ray observations probe the dense, X-ray--bright component of this circumgalactic medium (CGM), while the thermal Sunyaev--Zel'dovich (SZ) effect is also sensitive to lower-density gas through its integrated thermal pressure. We present the first spatially resolved joint X-ray and SZ analysis of a massive isolated disk galaxy, using deep \textit{XMM-Newton} observations of NGC~4594 from the X-raying the Accretion Reservoir Transferred to the ATmosphere Orbiting a Massive Spiral (XART-ATOMS) program, together with multiple independent \textit{Planck}-based SZ reconstructions. The measured SZ signal in the inner halo appears to exceed the value predicted from the X-ray-derived gas properties, even after accounting for major systematic uncertainties. Our preferred interpretation is that an additional lower-density CGM component contributes little to the observed soft X-ray emission but carries substantial thermal pressure. Under a simple two-phase model in approximate pressure equilibrium, the most probable X-ray-to-SZ ratio implies that the detected X-ray--emitting gas occupies only a small fraction of the halo volume, with a characteristic filling factor of $f_X \sim (4-6)\times10^{-3}$ near $r\sim50$~kpc. Such a small filling factor implies that single-phase X-ray analyses can underestimate the baryon mass contained in the hot CGM by a factor of $\sim2.5$, while the thermal energy content of the CGM is underestimated by about one order of magnitude. The non-detection of the additional component in the X-ray spectra requires it to be very hot and/or spatially extended, so weak in the soft X-ray emissions. These results favor the presence of a ``hidden'' X-ray--faint hotter halo component, while alternative interpretations, including a nonthermal contribution, are not excluded.
\end{abstract}

\keywords{
\uat{Circumgalactic medium}{1879} ---
\uat{X-ray astronomy}{1810} ---
\uat{Sunyaev-Zeldovich effect}{1654} ---
\uat{Galaxy evolution}{594} ---
\uat{Galaxy structure}{622}
}


\section{Introduction} \label{sec:introduction}

A substantial fraction of the baryons and feedback energy associated with massive galaxies is thought to reside in an extended circumgalactic medium (CGM) surrounding the galaxy (e.g., \citealt{Fukugita1998,McGaugh2010,Bregman2018}). In galaxies with halo masses comparable to or above that of the Milky Way, much of this material is expected to exist in a hot phase with temperatures near the virial temperature of the dark matter halo. This hot CGM is commonly studied through diffuse X-ray emission (e.g., \citealt{Anderson2013,LiJ2013a,LiJ2013b,LiJ2017}), ultraviolet or X-ray absorption-line spectroscopy (e.g., \citealt{Tumlinson2011,Nicastro2023}), and the thermal Sunyaev--Zel'dovich (SZ) effect (e.g., \citealt{Greco2015,Bregman2022}). However, each method probes different gas phases and physical quantities, and large uncertainties remain regarding the thermal, baryonic, and metal content of the hot halo.

X-ray observations directly probe the densest and brightest phase of the hot CGM and have revealed extended gaseous halos around nearby massive galaxies (e.g., \citealt{Anderson2013,LiJ2017}). However, X-ray emissivity depends strongly on gas density and metallicity, making X-ray measurements less sensitive to low-density diffuse gas and potentially biased toward denser structures (e.g., \citealt{LiJ2026}). The thermal SZ effect provides a complementary probe because it traces the integrated thermal pressure of ionized gas and is therefore more sensitive to diffuse low-density material (e.g., \citealt{Greco2015,Bregman2018,Bregman2022}). Most SZ studies of galaxy halos, however, rely on stacking analyses owing to the weakness of the signal from individual galaxies and the limited angular resolution of current instruments. Spatially resolved joint X-ray and SZ studies of individual galaxies remain rare.

Massive isolated disk galaxies provide favorable laboratories for studying the hot CGM \citep{LiJ2017,LiJ2018}. Their relatively quiescent environments reduce contamination from surrounding intragroup or intracluster gas, while their large halo masses help maintain stable hot atmospheres with long cooling times. Among nearby systems, the Sombrero galaxy (NGC~4594) is particularly well suited for such studies. NGC~4594 is the most massive isolated disk galaxy within $\sim 30$~Mpc, with a stellar mass several times that of the Milky Way and a halo mass of $M_{200} \sim 10^{13}~M_\odot$ \citep{Jiang2023}. Previous studies have detected both extended X-ray emission and a spatially resolved SZ signal around the galaxy (e.g., \citealt{Bregman2022}), making it one of the few nearby disk galaxies suitable for a direct comparison between the thermal pressure inferred from X-ray observations and that measured from the SZ effect.

In this letter, we present a joint X-ray and SZ study of the hot CGM surrounding NGC~4594. In \S\ref{sec:DataAnalysis}, we describe the \textit{XMM-Newton} and \textit{Planck} observations and the procedures used to derive the X-ray and SZ radial profiles. In \S\ref{sec:results}, we compare the observed SZ signal with that predicted from the X-ray-derived gas properties and quantify the discrepancy between them. In \S\ref{sec:Discussions}, we discuss the implications of this discrepancy for the multiphase structure of the hot CGM. Our major results and conclusions are summarized in \S\ref{sec:summary}.

\section{Data and Analysis} \label{sec:DataAnalysis}

\subsection{XMM-Newton observations} \label{subsec:XMM}

\begin{figure}
	\centering
	\includegraphics[width=0.45\textwidth]{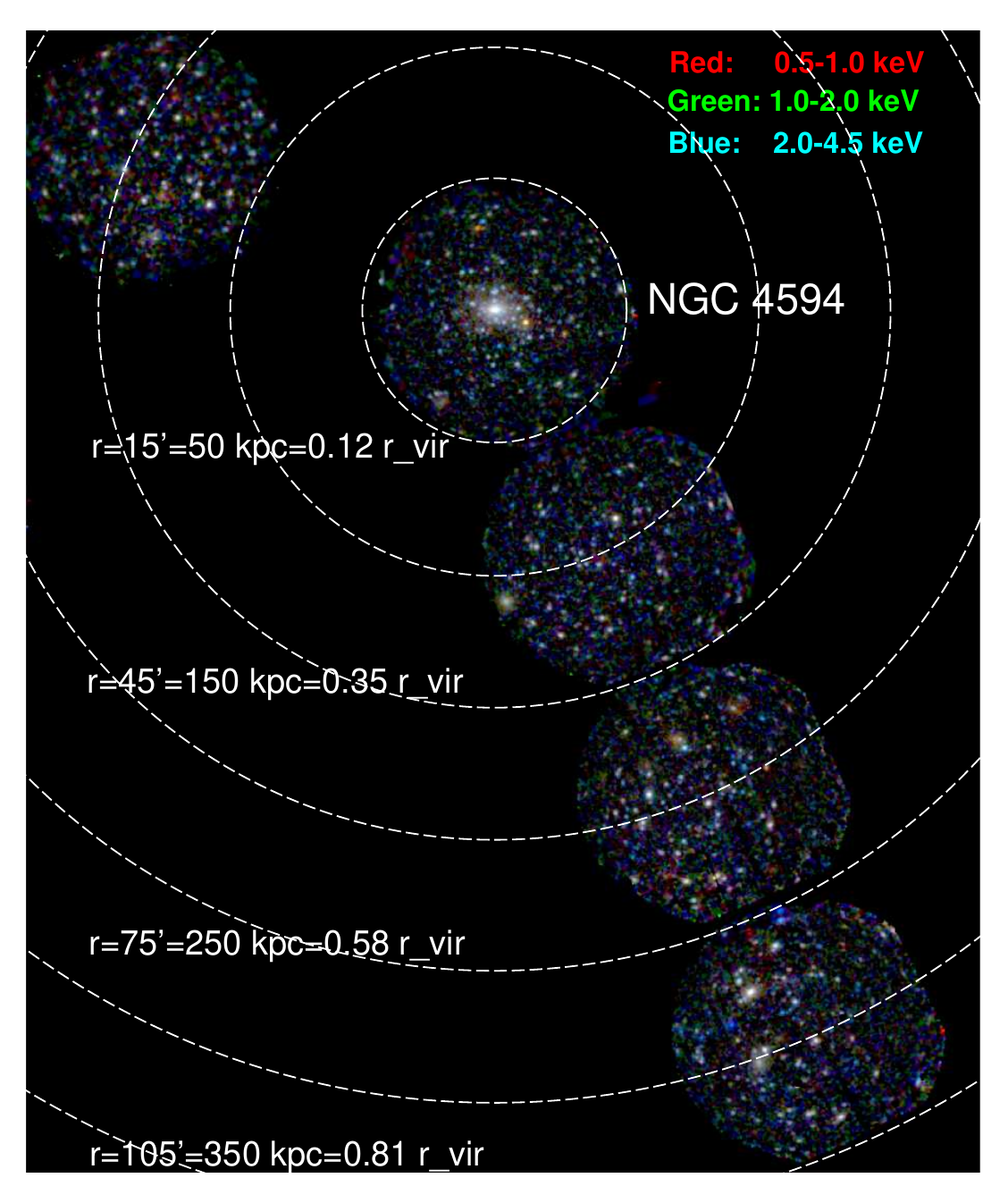}
	\caption{Tri-color X-ray view of NGC~4594 from \textit{XMM-Newton}.
		Composite image with \textbf{red} = 0.5–1\,keV, \textbf{green} = 1–2\,keV and \textbf{blue} = 2–4.5\,keV. Overlaid \textbf{dashed concentric annuli} mark reference galactocentric radii and illustrate that the XART-ATOMS programme provides nearly continuous coverage to $r\lesssim0.8\,r_{\rm vir}$. The observations used to construct this image are listed in Table~\ref{table:XMMdata}. North is up; East is left.}
	\label{fig:Xrayimages}
\end{figure}

We analyze \textit{XMM-Newton} observations of NGC~4594 obtained primarily from the X-raying the Accretion Reservoir Transferred to the ATmosphere Orbiting a Massive Spiral (XART-ATOMS) program, an AO-21 Large Program consisting of four contiguous pointings extending from the galaxy center to $r \approx 100^\prime$ ($\approx 330$~kpc). We additionally include one archival offset field located $\sim 45^\prime$ northeast of the galaxy to characterize azimuthal variations in the X-ray background (Fig.~\ref{fig:Xrayimages}). After flare filtering, the total effective MOS exposure time is $\sim 390$~ks (Table~\ref{table:XMMdata}).


\begin{figure*}
	\centering
    \includegraphics[width=0.325\textwidth]{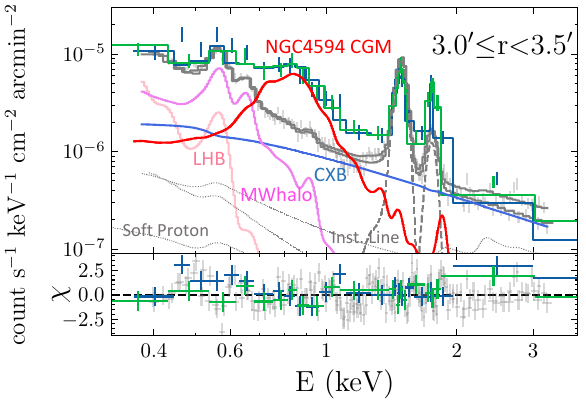}
    \includegraphics[width=0.325\textwidth]{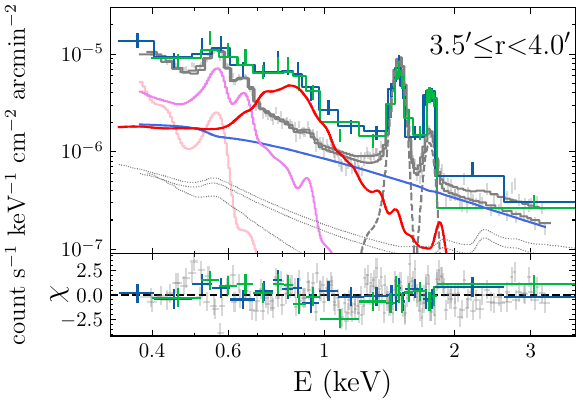}
    \includegraphics[width=0.325\textwidth]{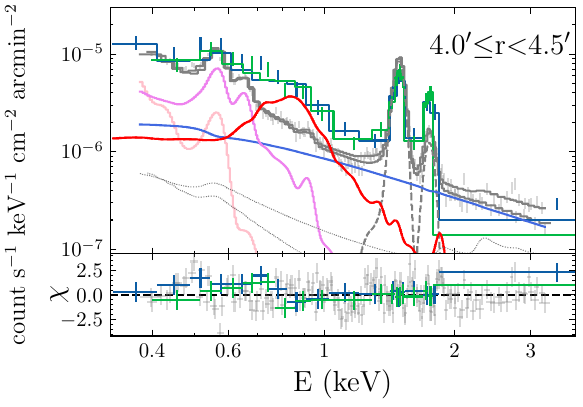}
    \includegraphics[width=0.325\textwidth]{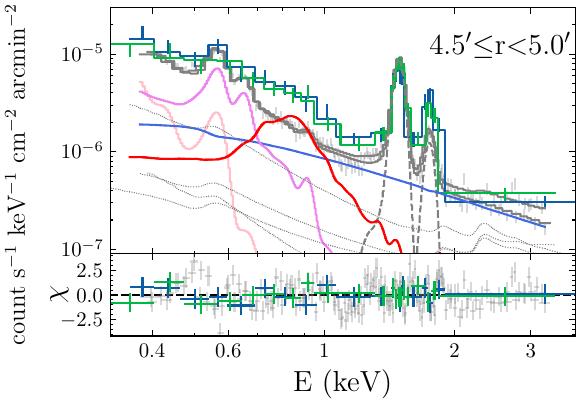}
    \includegraphics[width=0.325\textwidth]{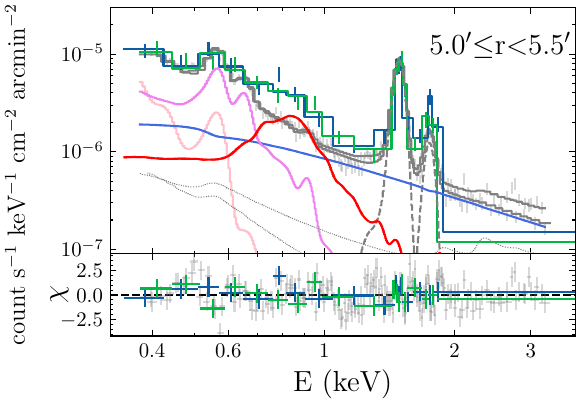}
    \includegraphics[width=0.325\textwidth]{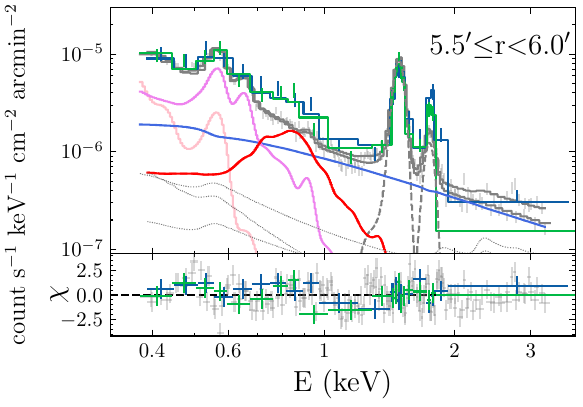}
    \includegraphics[width=0.325\textwidth]{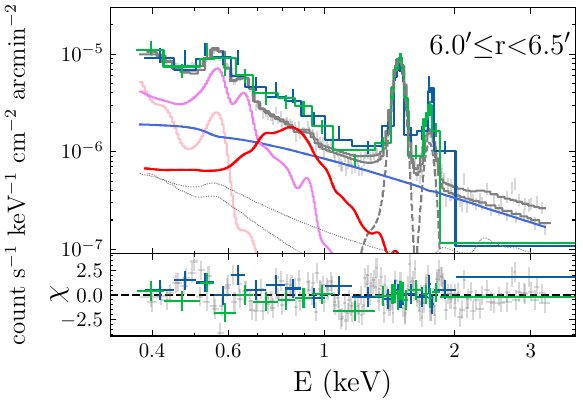}
    \includegraphics[width=0.325\textwidth]{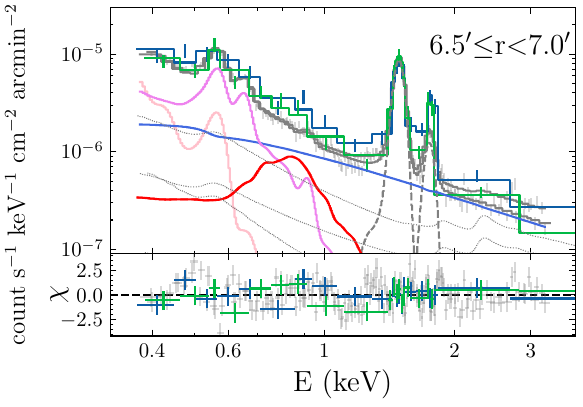}
    \includegraphics[width=0.325\textwidth]{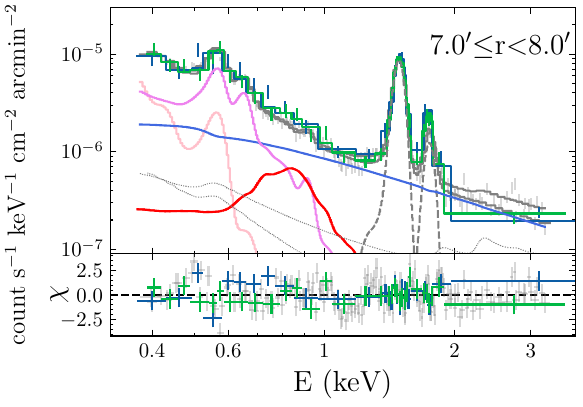}
    \includegraphics[width=0.325\textwidth]{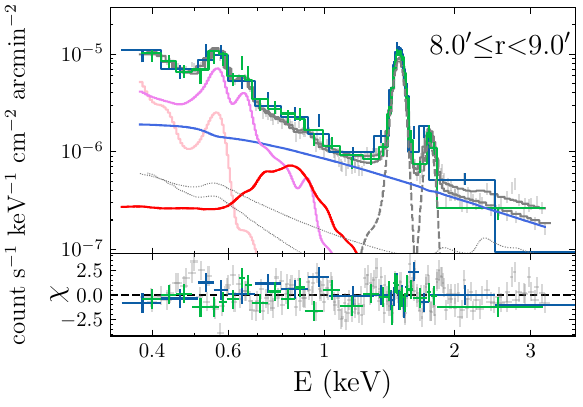}
    \includegraphics[width=0.325\textwidth]{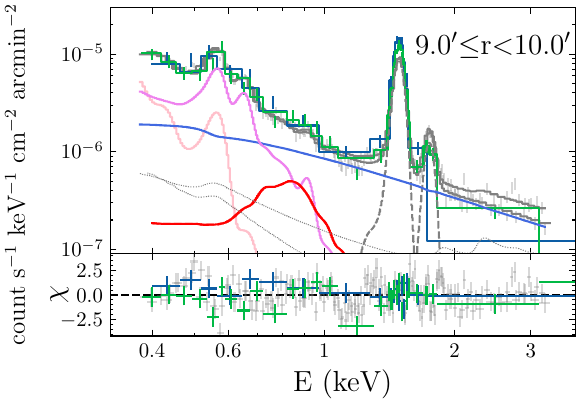}
    \includegraphics[width=0.325\textwidth]{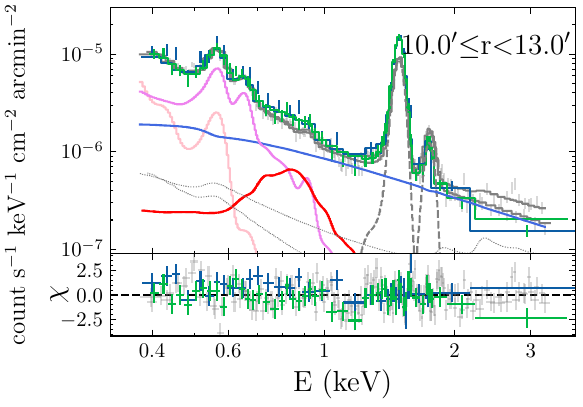}
    \caption{\textbf{\textit{XMM-Newton} spectra from concentric annuli of NGC~4594.}
    MOS1 (\textcolor{blue}{blue}) and MOS2 (\textcolor{green}{green}) spectra are shown with \(1\sigma\) error bars. Solid curves of the same colors are the total best-fit models for each annulus. For comparison, spectra from the outermost pointing (ObsID~0900170701; MOS1+MOS2; adopted as sky background) and their best-fit model are plotted in \textcolor{gray}{grey}. The source model is decomposed into its principal astrophysical components: Local Hot Bubble (LHB; \textcolor{pink}{pink}), Milky Way halo (\textcolor{magenta}{magenta}), cosmic X-ray background (CXB; \textcolor{blue}{blue}) and the hot CGM of NGC~4594 (\textcolor{red}{red}). Instrumental lines and the soft proton component are shown as \textcolor{gray}{grey} dashed and dotted curves, respectively. The lower panel shows residuals, \(\chi \equiv (\mathrm{data}-\mathrm{model})/\sigma\), where \(\sigma\) is the \(1\sigma\) uncertainty of each spectral bin.}
  \label{fig:XMMSpec}
\end{figure*}

We reduce the data using the \textit{XMM-Newton} Extended Source Analysis Software (\textsc{xmm-esas}; \citealt{Snowden2008}), following the procedures described in \citet{LiJ2017}. Spectra are extracted from concentric annuli centered on the nucleus of NGC~4594 (Fig.~\ref{fig:XMMSpec}). We use only the MOS1 and MOS2 data because the pn detector background is less stable for faint diffuse emission analyses. The spectra are fitted with an optically thin thermal plasma model representing the hot CGM together with astrophysical and instrumental background components, including the Local Hot Bubble (LHB), the Milky Way halo, the cosmic X-ray background, quiescent particle background, fluorescent instrumental lines, and residual soft proton contamination \citep{LiJ2017}.

\begin{figure}
	\centering
	\includegraphics[width=0.45\textwidth]{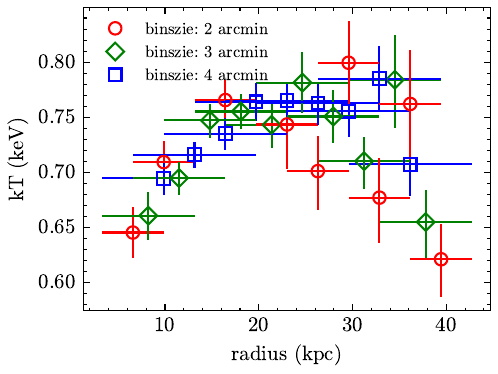}
	\caption{\textbf{Hot gas temperature profile of NGC~4594 from X-ray spectroscopy.}
    \textbf{Red}, \textbf{green} and \textbf{blue} points show profiles extracted with different annular bin widths. Vertical error bars indicate $1\sigma$ uncertainties from the spectral fits in each annulus.}
  \label{fig:XraykTprofile}
\end{figure}

The spectral fits yield radial profiles of the emission measure and temperature of the hot gas. Assuming spherical symmetry and a unit volume filling factor, we derive the electron density profile and fit it with a standard $\beta$ model:
\begin{equation}
n_e(r) = n_{e0}\left(1+\frac{r^2}{r_c^2}\right)^{-3\beta/2}.
\end{equation}
Because the temperature varies only mildly with radius (Fig.~\ref{fig:XraykTprofile}), we adopt a characteristic temperature of $kT \approx 0.75$~keV obtained from spectral analysis when computing the thermal pressure profile.


\begin{table*}
\begin{minipage}{\textwidth}
\caption{\textbf{\textit{XMM-Newton} observations of NGC~4594 used in this work.} $t_{\rm dur}$ denotes the raw on–source exposure, and $t_{\rm eff}$ is the effective exposure after background–flare filtering. Exposure times are given separately for MOS1 and MOS2. PN data are not used for calibration consistency.}
\begin{center}
\begin{tabular}{lccccccccccccc}
\hline
ObsID & PI & Object & RA & DEC & d & Start Date & $t_{\rm dur}$ (M1, M2) & $t_{\rm eff}$ (M1, M2) \\
 &  &  & deg & deg & arcmin &  & ks & ks \\
\hline
0900170101 & Li & NGC4594-1 & 189.9667 & -11.6417 & 2.14 & 2022-06-24 & 101.8, 100.7& 74.9, 78.5 \\
0900170501 & Li & NGC4594-2 & 189.7800 & -12.0981 & 31.24 & 2022-12-28 & 117.6, 117.6 & 75.9, 83.3 \\
0900170601 & Li & NGC4594-3 & 189.5912 & -12.5536 & 60.71 & 2023-01-03 & 121.9, 122.1 & 111.6, 112.6 \\
0900170701 & Li & NGC4594-4 & 189.4050 & -13.0129 & 90.34 & 2023-01-19 & 117.4, 117.7 & 100.1, 100.1 \\
0136950201 & Jansen & RXJ1242 & 190.6604 & -11.3225 & 42.94 & 2001-06-21 & 29.1, 29.0 & 28.2, 28.1 \\
\hline
\end{tabular}\label{table:XMMdata}
\end{center}
\end{minipage}
\end{table*}

The X-ray--predicted thermal SZ profile is obtained by projecting the thermal pressure along the line of sight:
\begin{equation}
\hat{y}_X(b) = \frac{\sigma_T}{m_ec^2}\int P_e(r)\,dl,
\end{equation}
where $b$ is the projected radius and the hat indicates the assumption that the X-ray--emitting gas uniformly fills the halo volume. 

We consider a few systematic uncertainties that affect the inferred hot CGM properties and their conversion to the SZ signal. We first investigate the influence of the assumed metallicity of the CGM component on the $EM$. At fixed $T=0.75$\,keV, fitting with metallicities $Z=\{1.0,\,0.3,\,0.1,\,0.03,\,0.01\}\,Z_\odot$ yields $EM$ changes (relative to $Z=0.3\,Z_\odot$) by factors of $0.3$, $1.0$, $2.3$, $3.8$, and $3.9$, respectively. This strong anti-correlation between $Z$ and the inferred density translates directly to the SZ prediction via \(\hat y_X \propto (EM)^{1/2}\). This is the major systematic uncertainty we considered in the following analysis.

We next assess how the assumed temperature affects the $EM$. The best-fit CGM temperature typically lies in the range $\sim0.6-0.85\rm~keV$ (Fig.~\ref{fig:XraykTprofile}). Holding metallicity fixed, $EM$ changes by only $\sim5\%$, $9\%$, $5\%$, and $17\%$ when $T$ is set to 0.6, 0.7, 0.8, and $0.9\rm~keV$ about the fiducial $T\approx0.75\rm~keV$. These shifts are smaller than the per-annulus statistical uncertainty on $EM$ ($\sim10\%-60\%$), implying that the X-ray–predicted SZ signal \((\hat y_X \propto EM^{1/2} k_{\rm B}T)\) is comparatively insensitive to the temperature assumption. We also tested a lognormal temperature distribution \citep{Cheng2021} against the single-temperature model; both fit the spectra equally well, and the resulting differences in $EM$ and in \(\hat y_X\) are below the statistical uncertainties. Given the narrow energy band in which the CGM dominates the emission (Fig.~\ref{fig:XMMSpec}), details of the temperature distribution do not significantly affect our systematic error budget.

Background systematics include astrophysical components (CXB, LHB, MW halo, eROSITA bubbles, possible solar-wind charge exchange SWCX, and residual point/extended sources) and non-astrophysical components (QPB and soft protons). Our hot CGM signal is concentrated in $0.7-1.0$\,keV, where contributions from the LHB ($kT\sim0.1$\,keV) and the MW halo/eROSITA bubbles ($kT\sim0.2$\,keV) are small, and SWCX is negligible in all of the observations used for CGM extraction. To empirically capture background fluctuations, we analyzed the spectra extracted from various source-free regions at $15^{\prime}<r<105^{\prime}$ across all available \textit{XMM-Newton} fields; the $0.7-1.0$\,keV flux is $(5.6\pm0.6)\times10^{-7}$\,erg\,s$^{-1}$\,cm$^{-2}$ (a $\sim10\%$ variation). Because the CXB dominates the background in this band (Fig.~\ref{fig:XMMSpec}), we propagate a $\pm10\%$ uncertainty on the CXB normalization. The induced change in $EM$ depends strongly on radius: $\sim(3-10)\%$ in the inner annuli ($r=3^{\prime}-7^{\prime}$), rising to $\sim(10–30)\%$ at $r=7^{\prime}-10^{\prime}$, and $\sim(30-40)\%$ at $r=10^{\prime}-13^{\prime}$ as the CGM surface brightness declines (difficult to estimate at $r=13^{\prime}-15^{\prime}$ due to the high instrument background at the boundary of the first FOV; Fig.~\ref{fig:Xrayimages}). Beyond $r\gtrsim15^{\prime}$, a $\sim10\%$ CXB fluctuation typically reduces the CGM to an upper limit constraint.

In conclusion, within the radii most relevant for this work ($r\sim10^{\prime}$--$20^{\prime}$; \S\ref{sec:results}), metallicity is the dominant systematic for the X-ray-predicted SZ profile; temperature and background fluctuations are less important. At larger radii, background fluctuations increasingly limit the $EM$ measurement and thus $\hat y_X$. In the quantitative comparison between the X-ray-predicted and directly observed SZ profiles we therefore include metallicity-driven systematics for the inner regions; uncertainties at larger radii should be correspondingly larger due to the background fluctuation but do not affect our conclusions.

\subsection{SZ observations} \label{subsec:SZ}

\begin{figure}
	\centering
    \includegraphics[width=0.45\textwidth]{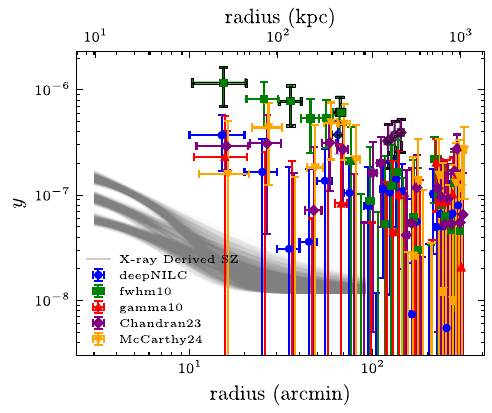} 
	\caption{Radial SZ profiles of NGC~4594: measurements versus X-ray–predicted curves.
    \textbf{Grey curves} are the X-ray–predicted SZ profiles derived from the hot gas parameters, evaluated for three assumed metallicities: $0.1\,Z_\odot$ (top cluster), $0.3\,Z_\odot$ (middle) and $1.0\,Z_\odot$ (bottom). \textbf{Coloured symbols} are the measured SZ profiles from five independent \textit{Planck}-based maps: \texttt{Chandran23} (\textcolor{red}{red triangles}), \texttt{McCarthy24} (\textcolor{violet}{purple diamonds}), \texttt{fwhm10} (\textcolor{green!60!black}{green squares}), \texttt{gamma10} (\textcolor{orange}{orange inverted triangles}) and \texttt{deepNILC} (\textcolor{blue}{blue circles}). To aid visibility, symbols within each radial bin are given a slight horizontal offset; the \texttt{deepNILC} point marks the bin center.}
	\label{fig:XraySZprofiles}
\end{figure}

We extract the thermal SZ signal from different annuli around NGC~4594 (Fig.~\ref{fig:XraySZprofiles}), excluding the inner $\sim33\rm~kpc$ ($r\approx10^\prime\approx 0.08 r_{\rm vir}$) region that defines the optical galaxy and might have contamination from thermal dust emission \citep{Jiang2023}. Potential contamination from a central radio source and $\sim10$\,kpc radio lobes \citep{Yang2024} are also mitigated. 

SZ signals from galaxies are generally much weaker than from a galaxy cluster, and for weak signals, the result depends on the SZ map that is used.  There are a variety of choices when making such a map, but we restrict our choices to those that use the most recent data release of \emph{Planck} data (2020; Data Release~4).  The \emph{Planck} data have a Low Frequency Instrument (LFI) at 30, 44, and 70~GHz and a High Frequency Instrument (HFI) at 100, 143, 217, 353, 545, and 857~GHz for a total of nine channels.  One can also include data from the \emph{WMAP} observatory, which uses five frequency bands: 23~GHz, 33~GHz, 41~GHz, 61~GHz, and 94~GHz.  The signal and resolution of the \emph{Planck} data are superior to those of \emph{WMAP}, but \emph{WMAP} is very well calibrated and there are differences between the data from the two observatories, where they overlap, indicating weak systematic differences.  Such differences can affect the resulting SZ map, so using data from both observatories has some benefit.

The choice of the bands to be used depends on what one seeks to study, and for thermal SZ signals, the null point of the SZ signal is at 217~GHz, with a negative signal below that frequency and a positive signal above.  So the frequencies around 217~GHz contribute the most to the net signal, but one must also use other frequency bands to characterize signals that contaminate the SZ signal, such as thermal dust emission as well as the CMB signal. Of the possible channels, the 857~GHz band has the largest noise and little SZ contribution, so it is often not used in map construction.  The lowest frequency signals, such as at 20–33~GHz, have the poorest spatial resolution and may only provide weak constraints on the SZ signal, so some researchers do not use these channels in their SZ maps.

Another consideration in map making is the window functions used in the harmonic analysis and the shape of filters in the spatial analysis.  One can choose the various functions to tailor the resulting map for point-source detection or for extended emission, or to minimize the contribution of some known contaminant, such as far-infrared emission from distant galaxies (e.g., \citealt{McCarthy2024}).  There is also a well-known bias between the root-mean-square of the resulting map and the independence of signals on various spatial scales.  One makes choices depending on the scientific question being asked and the degree of bias that one is willing to accept.  This leads to a variety of different possible maps, the limitations of which are not always explored in depth.  It is striking that, for weak signals, there can be significant differences in the $y$ values between various maps.

We use five SZ maps constructed with different component-separation techniques and data combinations: \texttt{McCarthy24} \citep{McCarthy2024}, \texttt{Chandran23} \citep{Chandran2023} based on \emph{Planck} PR4 data, and three maps developed by us that combine \emph{Planck} and \emph{WMAP} data, namely \texttt{fwhm10} \citep{Pratt2021}, \texttt{gamma10} \citep{Bregman2022}, and the deep-learning-based \texttt{deepNILC} \citep{Pratt2024}. These maps differ primarily in their implementation of internal linear combination (ILC) or needlet ILC (NILC) techniques, the adopted frequency channels, and the degree of spatial and spectral localization. In particular, the \texttt{Chandran23} and \texttt{McCarthy24} maps use updated PR4 data with improved calibration and reduced large-scale systematics, including striping and dust contamination \citep{Chandran2023}, while the three maps constructed by our group additionally incorporate \emph{WMAP} data to improve large-scale stability and cross-calibration \citep{Bregman2022,Pratt2021,Pratt2024}. The \texttt{fwhm10} and \texttt{gamma10} maps adopt different spatial filtering and localization schemes within the NILC framework, whereas \texttt{deepNILC} combines multiple NILC realizations with varying localization parameters through a neural network to mitigate the bias--variance trade-off inherent in standard NILC methods and to improve the recovery of weak, extended SZ signals, as demonstrated in simulations \citep{Pratt2024}.

These methodological differences lead to distinct systematic uncertainties. The primary sources include (i) residual foreground contamination, particularly from Galactic dust and the cosmic infrared background (CIB), which depends sensitively on the frequency weighting and component-separation assumptions \citep{McCarthy2024}; (ii) the treatment of spatial localization, which controls the balance between noise suppression and signal bias in NILC-based maps \citep{Pratt2024}; and (iii) instrumental noise and large-scale systematics, such as scan-induced striping, which are reduced but not eliminated in the latest PR4 products \citep{Chandran2023}. As a result, the maps are highly correlated on large angular scales ($\gtrsim 1$--$2^\circ$), where the signal is dominated by common foreground residuals and large-scale modes, but show significant differences on smaller scales, where the SZ signal is extracted through different filtering and contamination-mitigation strategies. These small-scale differences dominate the uncertainties in the radial SZ profiles of NGC~4594 (Fig.~\ref{fig:XraySZprofiles}).

To minimize contamination, we apply a common mask to all maps that excludes known galaxy clusters, groups, radio sources, and other foreground or background objects. We further inspect each map individually to remove residual artifacts not captured by the automated masking. The innermost region ($r<10^\prime$), corresponding to the optical galaxy and associated radio structures \citep{Yang2024}, is excluded from the analysis to avoid contamination from non-SZ emission.


\begin{figure}
    \centering
    \includegraphics[width=0.45\textwidth]{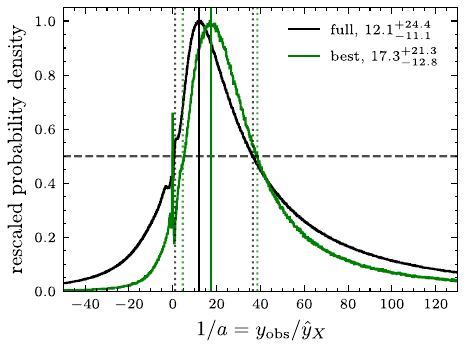} 
    \caption{The rescaled unit-peak PDFs of the ratio $1/a\equiv y_{\text{obs}}/\hat{y}_{X}$ in the $r=10^\prime-20^\prime$ ($r\approx50\rm~kpc$) annulus. Two datasets are shown: ``full'' (\textbf{black}) uses all five SZ maps and X-ray–predicted profiles for \(Z=\{0.1,0.3,1.0\}\,Z_\odot\); ``best'' (\textbf{green}) uses only the most reliable \texttt{deepNILC} SZ map and the \(Z=0.3\,Z_\odot\) X-ray prediction. The peak and FWHM for each dataset are listed in the panel legends. The \textbf{horizontal dashed line} marks the half-maximum level (identical across datasets due to the unit-peak normalization), and the \textbf{vertical dotted lines} indicate the FWHM bounds for each PDF.}\label{fig:XraySZHist}
\end{figure}

\section{Results} \label{sec:results}

Using the radial density and temperature profiles derived from the \textit{XMM-Newton} observations, we compute the thermal SZ profile expected from the X-ray--emitting hot gas by projecting the inferred thermal pressure profile along the line of sight. Fig.~\ref{fig:XraySZprofiles} compares the observed SZ radial profiles ($y_{\rm obs}$) with the X-ray--predicted SZ profiles ($\hat{y}_X$) assuming different metallicities.

Across a wide radial range, the observed SZ signal systematically exceeds the value predicted from the X-ray-derived gas properties. This discrepancy persists across all analyzed SZ maps and cannot be removed by adopting different plausible metallicities for the hot gas (or considering other systematic uncertainties). To quantify the discrepancy between the X-ray--predicted and observed SZ signals, we construct probability density functions (PDFs) for the ratio $1/a\equiv y_{\rm obs}/\hat{y}_X$, using Monte Carlo sampling (Fig.~\ref{fig:XraySZHist}). The uncertainty in the observed SZ signal is estimated from both the statistical uncertainties within each SZ map and the scatter among the different SZ reconstructions. For the X-ray prediction, uncertainties are estimated from the fitted $\beta$-model parameters together with the assumed metallicity range. We consider two representative cases: a ``full'' sample including all SZ maps and metallicity assumptions, and a ``best'' sample using only our preferred \texttt{deepNILC} map \citep{Pratt2024} together with a relatively low fiducial metallicity of $0.3~Z_\odot$ \citep{HodgesKluck2018}.

Fig.~\ref{fig:XraySZHist} shows the resulting PDFs for the first annulus of the observed SZ profile at $r\approx50\rm~kpc$ ($r=10^\prime$--$20^\prime$). Because the PDFs are significantly asymmetric, we characterize them using the peak value and the full width at half maximum (FWHM). For comparison, under a symmetric Gaussian distribution, the FWHM corresponds to $\approx\pm1.177~\sigma$, or $\sim76\%$ confidence level. In this bin, the observed SZ signal exceeds the X-ray prediction with probabilities of $\sim 86\%$ for the ``full'' sample and $\sim 94\%$ for the ``best'' sample. The corresponding ratios between the X-ray--predicted and observed SZ signals are $a \approx 0.08$ (0.06) for the ``full'' (``best'') sample at $r \approx 50$~kpc. Even when integrating over the full radial range covered by the \textit{XMM-Newton} observations ($r \lesssim 330$~kpc), the probability that the observed SZ signal exceeds the X-ray prediction remains at $\sim 63\%$ ($\sim 68\%$) for the ``full'' (``best'') sample. Although both PDFs peak well above $y_{\rm obs}/\hat{y}_X=1$, the equality between the observed and X-ray--predicted SZ signals is not excluded at very high statistical significance.  The filling factor and bias factors discussed below in \S\ref{sec:Discussions} are therefore based on the most probable values of the ratio and would be substantially less extreme if the true ratio were closer to unity.

\section{Discussion and Prospects} \label{sec:Discussions}

We present the first solid and direct spatially resolved evidence for the discrepancy between the X-ray--predicted and observed SZ signals within an individual massive isolated disk galaxy. Although similar X-ray--SZ discrepancies have previously been suggested \citep{Bregman2018,Zhang2024}, the results in those works are based on stacked analyses of massive galaxies and large-scale structures with large and unquantified systematic uncertainties. The X-ray--SZ discrepancy strongly suggests that the CGM is two phases, one produces both X-ray and SZ signals, while the other produces additional SZ signals but negligible X-ray emissions. 

In addition to the X-ray--emitting phase, a lower-temperature phase is unlikely to explain the observed SZ excess. Gas with temperatures below $\sim 10^6$~K cools efficiently and would require unrealistically large pressure support to produce a substantial SZ signal while remaining X-ray faint. A more plausible explanation is a hotter and more diffuse CGM phase with lower density and therefore much weaker X-ray emissivity. Such a phase could naturally arise from thermalized Type Ia supernova feedback or episodic AGN heating operating in the bulge and halo of massive galaxies \citep{Tang2009,Tang2010,LiJ2026}. Owing to its low density, this component would contribute little to the observed X-ray emission while still carrying substantial thermal pressure detectable through the SZ effect.

To characterize the structure of the two-phase hot CGM, we consider a toy model in which the X-ray--emitting phase has a volume filling factor $f_X$, while the hotter X-ray--faint phase has a filling factor $1-f_X$. The X-ray emission measure $EM$ for a shell of volume $V$ can be written as:
\begin{equation}\label{eq:EMfX}
    EM = \int n_{\rm e,X}n_{\rm H,X} \, dV \;\;\approx\;\; n_{e,X}^2 \, f_X V ,
\end{equation}
which yields:
\begin{equation}\label{eq:nenefit}
    n_{e,X} = \frac{n_{e,\mathrm{fit}}}{\sqrt{f_X}}, \qquad
    n_{e,\mathrm{fit}} \equiv \sqrt{\frac{EM}{V}},
\end{equation}
where $n_{e,\mathrm{fit}}$ is the electron density one would infer under the assumption of $f_X=1$.

The SZ contribution from the X-ray--emitting phase is then:
\begin{equation}\label{eq:SZXray}
    y_X = \frac{\sigma_T}{m_e c^2} \, n_{e,X} kT_X \, f_X \ell
         = \hat{y}_X \, \sqrt{f_X},
\end{equation}
where $\ell$ is the effective line-of-sight depth of the shell, $T_X$ is the temperature of the X-ray emitting hot gas, and
\begin{equation}
    \hat{y}_X \equiv \frac{\sigma_T}{m_e c^2} \, n_{e,\mathrm{fit}} kT_X \, \ell
\end{equation}
is the naive SZ signal one would compute from X-ray data assuming $f_X=1$.

Assuming a pressure ratio of the two phases $R_{Pth} \equiv P_h/P_X$ ($h$ denotes the hotter phase), the observed SZ signal is then:
\begin{equation}
     y_{\mathrm{obs}} = \frac{\sigma_T}{m_e c^2} \, \ell P_X \left[f_X + R_{Pth}(1-f_X)\right].
\end{equation}

We can then express the ratios $y_X/y_{\mathrm{obs}}$ and $a \equiv \hat{y}_X/y_{\mathrm{obs}}$ as:
\begin{equation}
    \frac{y_X}{y_{\mathrm{obs}}} = \frac{f_X}{f_X + R_{Pth}(1-f_X)},
\end{equation}
and:
\begin{equation}\label{eq:fXequ}
    a(1-R_{Pth})f_X - \sqrt{f_X} + aR_{Pth} = 0,
\end{equation}
which can be solved for $f_X$:
\begin{equation}\label{eq:fXnonPbalance}
f_X \;=\; [\frac{1 - \sqrt{\,1 - 4a^2 R_{Pth}(1-R_{Pth})\,}}{2a(1-R_{Pth})} ]^2.
\end{equation}
Eq.~\ref{eq:fXnonPbalance} gives an estimate of $f_X$ from the observable $a$. Under pressure balance ($R_{Pth}=1$), this reduces to the simple expression: $f_X = a^2$. Adopting the observed $a\approx0.08$ for the ``full'' sample at $r \approx 50$~kpc (\S\ref{sec:results}), we infer a characteristic filling factor of $f_X \sim 6\times10^{-3}$, indicating that the X-ray--emitting gas occupies only a small fraction of the halo volume.

We further compute the multiplicative \emph{mass bias factor}, in order to quantify the bias in estimating the total mass of the CGM based only on the X-ray data and a single phase assumption ($M_{\mathrm{X,only}}$):
\begin{equation}\label{eq:massfactor}
\;\mathcal{U}_M \;\equiv\; \frac{M_{\rm true}}{M_{\mathrm{X,only}}}
\;=\; \frac{ f_X + R_{Pth}\,\vartheta^{-1}\,(1-f_X) }{ \sqrt{f_X} } \;,
\end{equation}
where $M_{\rm true}$ is the true mass of the two-phase hot CGM, and $\vartheta\equiv T_h/T_X$ is the temperature ratio of the two phases. In reasonable parameter spaces of $R_{Pth}$, $\vartheta$, and $f_X$ (e.g., $R_{Pth}=1$, $\vartheta=5$, $f_X = 6\times10^{-3}$), the mass bias factor $\mathcal{U}_M\sim2.5$. Therefore, under the two-phase hot CGM scenario, assuming a single gas phase with $f_X=1$ could significantly underestimate the baryon mass contained in the hot CGM, helping to account for part of the ``missing baryons'' (e.g., \citealt{Bregman2022}).

Similarly, we also compute the \emph{energy bias factor}:
\begin{equation}\label{eq:energyfactor}
\mathcal{U}_E \;\equiv\; \frac{E_{\rm true}}{E_{\rm X,only}}
\;=\; \frac{f_X + R_{Pth}(1-f_X)}{\sqrt{f_X}}
\end{equation}
Under pressure balance ($R_{\rm Pth}=1$), \(\displaystyle \mathcal{U}_E=1/\sqrt{f_X}\ge 1\). Therefore, with the measured $f_X \sim 6\times10^{-3}$, incorrectly assuming $f_X=1$ could underestimate the total thermal energy in the CGM by an order of magnitude.

Similarly, the two-phase model also predicts a bias in the X-ray luminosity.  Let $\Lambda_X$ and $\Lambda_h$ denote the band-limited emissivities of the X-ray--emitting phase and the hotter X-ray--faint phase, respectively.  The true luminosity of the two-phase gas is: 
\begin{equation}
L_{\rm true} = \Lambda_X n_{e,X}^2 f_X V + \Lambda_h n_{e,h}^2 (1-f_X)V .
\end{equation}
Using
\begin{equation}
\frac{n_{e,h}}{n_{e,X}} = \frac{P_h/P_X}{T_h/T_X} = \frac{R_{\rm Pth}}{\vartheta},
\end{equation}
the luminosity bias factor is:
\begin{equation}\label{eq:luminosityfactor}
\mathcal{U}_L \equiv \frac{L_{\rm true}}{L_{X,{\rm only}}} = 1 + \frac{\Lambda_h}{\Lambda_X} \frac{R_{\rm Pth}^2}{\vartheta^2} \frac{1-f_X}{f_X}.
\end{equation}
If the emissivity ratio is approximated as $\Lambda_h/\Lambda_X \approx 1$ and the two phases are under pressure balance ($R_{\rm Pth}=1$), we can estimate the luminosity bias factor as $\mathcal{U}_L \approx 7-8$, adopting the measured $f_X\sim 6\times 10^{-3}$, and for a representative temperature contrast of $\vartheta=5$, so the hot component should have a luminosity a few times of the X-ray detected component. This is apparently in contrast to the non-detection of the hotter component, unless the emissivities of the two components are significantly different in the given band.


\begin{figure}
    \centering
    \includegraphics[width=0.45\textwidth]{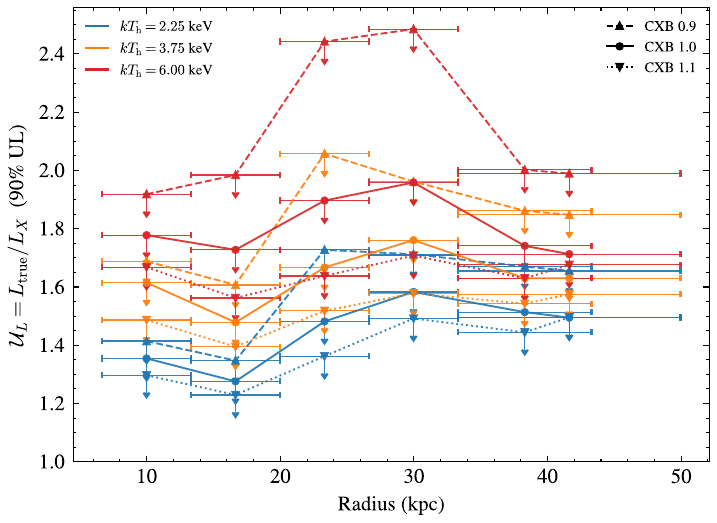} 
    \caption{Upper limits on the luminosity bias factor $\mathcal{U}_L$ (Eq.~\ref{eq:luminosityfactor}) as a function of galactocentric radius. The limits are computed in the $0.5$--$10$~keV band by adding a second thermal plasma component to the \emph{XMM-Newton} spectral fits.  The temperature of the detected soft X-ray component is fixed at $kT_X=0.75$~keV, while the temperature of the hotter component is fixed at $kT_h=2.25$, $3.75$, or $6.00$~keV, corresponding to $\vartheta=T_h/T_X=3$, 5, and 8 (plotted in different colors). All abundances are fixed at $0.3~Z_\odot$.  Different line styles show the effect of varying the CXB normalization by $\pm 10\%$ around the fiducial value (\S\ref{subsec:XMM}).  The plotted values are 90\% confidence upper limits on the total luminosity bias allowed by the X-ray spectra.}\label{fig:FluxRatioULvsR}
\end{figure}

To test whether such a hotter phase should have been detected in the X-ray spectra, we add a second thermal plasma component to the \emph{XMM-Newton} spectral fits in different annuli. The temperature of the detected soft X-ray--emitting phase is fixed at $kT_X=0.75$~keV.  We then fix the temperature of the additional component at $kT_h=2.25$, $3.75$, and $6.00$~keV, corresponding to temperature contrasts of $\vartheta=T_h/T_X=3$, 5, and 8. The metallicities of both components are fixed at $0.3~Z_\odot$.  Following the background tests described in \S\ref{subsec:XMM}, we repeat the fits after rescaling the CXB normalization by $\pm 10\%$ to bracket the dominant background systematic uncertainty.  For each assumed $T_h$, we derive the 90\% confidence upper limit on the normalization of the hotter component and convert it to an upper limit on the luminosity bias factor $\mathcal{U}_L$ in the $0.5$--$10$~keV band. The results are shown in Fig.~\ref{fig:FluxRatioULvsR}.

The X-ray spectra typically allow only $\mathcal{U}_L\lesssim 1.2$--$2.5$, depending on radius, assumed $T_h$, and the adopted CXB normalization.  These limits are below the simple two-phase expectation $\mathcal{U}_L\sim 7$--8 obtained above for $f_X\simeq 6\times10^{-3}$, $R_{\rm Pth}=1$, $\vartheta=5$, and $\Lambda_h/\Lambda_X\simeq1$.  This tension shows that a hotter component with approximately the same spatial distribution as the detected X-ray--emitting gas and substantial emissivity in the \emph{XMM-Newton} band is difficult to hide.  There are, however, important limitations. At the radii most relevant for the X-ray--SZ comparison ($r\sim10^\prime$--$20^\prime$), the X-ray surface brightness is already low and the constraint on an additional hot component is sensitive to various systematics (\S\ref{subsec:XMM}).  Conversely, at smaller radii where the X-ray spectra have better statistics, the \emph{Planck}-based SZ measurements have insufficient angular resolution to isolate the same physical region, and the temperature of the detected gas may also decline (Fig.~\ref{fig:XraykTprofile}). Thus the comparison between the X-ray upper limits and the SZ-inferred luminosity bias should be regarded as a useful consistency test rather than a one-to-one measurement at the same spatial scale.

A possible alternative contribution to the apparent SZ excess is a population of relativistic electrons associated with fossil radio lobes.  The known central radio source and the observed $\sim10$~kpc radio lobes are contained within the excluded $r<10^\prime$ region \citep{Yang2024}, but larger and fainter fossil lobes extending to tens of kiloparsecs cannot presently be ruled out.  Such lobes could, in principle, produce a nonthermal SZ distortion if their cosmic-ray pressure and line-of-sight extent are sufficiently large. However, no extended fossil lobes are currently detected around NGC~4594, and the conversion of a nonthermal spectral distortion into the thermal Compton-$y$ values reported by the \textit{Planck}-based maps is model dependent, particularly on the low-energy cutoff and spectral index of the electron population \citep{Colafrancesco2008}.  Deep spatially resolved low-frequency radio imaging will be required to determine whether fossil radio lobes contribute significantly to the measured signal (e.g., \citealt{Heald2022,Heesen2024}).

If the SZ excess is instead primarily produced by an additional thermal gas component, its non-detection in the X-ray spectra can be reconciled with its large SZ contribution if the phase is very hot, sufficiently extended, or has a much lower band-limite emissivity than assumed in the simple estimate above. Increasing $T_h$ reduces the gas density required to supply a given thermal pressure and shifts a larger fraction of the emission outside the soft X-ray band where the detected $kT_X\simeq0.75$~keV component is best constrained.  This interpretation, however, introduces two related physical difficulties.  First, if $kT_h\gtrsim$ several keV, the gas temperature is far above the characteristic virial temperature of the Sombrero halo ($kT_{\rm vir}\sim0.3$--$0.4$~keV for $M_{200}\sim10^{13}~M_\odot$; \citealt{Jiang2023}), and is therefore difficult to confine within the gravitational potential of the galaxy alone.  This confinement problem could be alleviated if part of the SZ signal arises from a more extended, group-scale atmosphere with a longer path length and a deeper effective potential, although NGC~4594 has no major nearby companions and is usually regarded as a relatively isolated massive disk galaxy.  Second, the required temperature can approach or exceed the specific energy expected from Type~Ia supernova ejecta (e.g., \citealt{LiJ2009,LiJ2011,LiJ2026,Tang2009,Tang2010}), making Type~Ia supernova heating alone insufficient. Additional energy input from AGN activity may therefore be required.  In this respect, the recently discovered $\sim10$~kpc radio lobes in NGC~4594 \citep{Yang2024} provide evidence for past nuclear activity that could contribute to heating, pressurizing, or lifting hot gas in the halo. Thus, while a very hot X-ray--faint phase remains a possible explanation for the X-ray--SZ discrepancy, its confinement and heating likely require a combination of an extended gravitational potential and non-stellar energy input.

\section{Summary and Prospects}\label{sec:summary}

We presented a joint X-ray and thermal SZ analysis of the hot CGM around the massive isolated disk galaxy NGC~4594. Using deep \textit{XMM-Newton} observations from the XART-ATOMS program, we derived the density and temperature profiles of the X-ray--emitting gas and used them to predict the corresponding Compton-$y$ profile. We then compared this X-ray--predicted SZ signal with measurements from multiple \textit{Planck}-based SZ reconstructions. The observed SZ signal in the inner halo, especially at $r\approx 50$~kpc, exceeds the value expected from the detected X-ray--emitting gas, and this discrepancy persists under different SZ map choices and plausible metallicity assumptions.

This X-ray--SZ discrepancy indicates that the X-ray--emitting phase alone does not account for the full thermal pressure of the halo. Our preferred interpretation is a simple two-phase model, in which the detected $kT_X\simeq 0.75$~keV gas occupies only a small fraction of the volume, while an additional hotter and more diffuse component contributes substantially to the SZ signal but remains weak in soft X-ray emission. The filling factor of the soft X-ray emitting gas implied by the most probable ratio $a\equiv \hat{y}_X/y_{\rm obs}$ is $f_X\simeq a^2\sim 4$--$6\times10^{-3}$ for the inner halo, under pressure equilibrium of the two hot gas components. Such a small filling factor implies that single-phase X-ray analyses can strongly underestimate both the baryon mass and the thermal energy content of the CGM. At the same time, the non-detection of the hotter component in the X-ray spectra places important constraints on its temperature, emissivity, spatial distribution, and physical origin. If the hidden component is thermal plasma in collisional ionization equilibrium, it may need to be very hot or spatially extended, possibly requiring a group-scale atmosphere, additional AGN heating, or both. A smaller intrinsic discrepancy, a scenario in which the X-ray and SZ signals arise largely from the same gas, or a nonthermal contribution from an extended relativistic-electron population cannot currently be excluded.

Next-generation observations and simulations will be essential for advancing studies of the hot CGM. Upcoming SZ facilities such as CMB-S4 will deliver finer angular resolution, sharpening measurements for distant clusters but remain challenged by the faint, very extended SZ signal from nearby galaxy halos (e.g., \citealt{Abazajian2019}). On the other hand, future X-ray missions with high-resolution microcalorimeters will separate emission lines from the continuum, provide robust metallicities, and reveal very hot gas through diagnostic line ratios and clean continuum shapes (e.g., \citealt{Bregman2023}). Narrow-band X-ray imaging of individual lines will further suppress sky and particle background, enabling deeper maps of hot gas to large galactocentric radii (e.g., \citealt{LiJ2018,LiJ2020}). Equally important are improved hydrodynamical simulations that resolve the multiphase structure of the halo and provide forward models to interpret joint X-ray and SZ observations.

\begin{acknowledgments}
We thank the anonymous referee for constructive comments that helped improve the clarity and interpretation of this paper. J.T.L. acknowledge the financial support from the National Science Foundation of China (NSFC) through the grants 12321003 and 12273111, from the science research grants from the China Manned Space Program with grant no. CMS-CSST-2025-A10 and CMS-CSST-2025-A04, and from the Jiangsu Innovation and Entrepreneurship Talent Team Program through grant JSSCTD202436.
\end{acknowledgments}

\bibliography{XARTATOMS_AAS}{}
\bibliographystyle{aasjournalv7}



\end{document}